\documentclass[twocolumn]{jpsj2} 
\usepackage{bm}

\title{Enhanced spin Hall effect in semiconductor heterostructures with
artificial potential}

\author{Mikio
\textsc{Eto}\thanks{E-mail address: eto@rk.phys.keio.ac.jp}
and Tomohiro \textsc{Yokoyama}}

\inst{Faculty of Science and Technology, Keio University,
3-14-1 Hiyoshi, Kohoku-ku, Yokohama 223-8522, Japan
\\
{\rm (received April 23, 2009)}}

\abst{We theoretically investigate an extrinsic spin Hall effect
(SHE) in semiconductor heterostructures due to the scattering by an
artificial potential created by antidot, STM tip, etc. The
potential is electrically tunable.
First, we formulate the SHE in terms of phase shifts in the
partial wave expansion for two-dimensional electron gas.
The effect is significantly enhanced by the resonant scattering
when the attractive potential is properly tuned.
Second, we examine a three-terminal device including an antidot,
which possibly produces a spin current with  polarization of
more than 50\%.
}

\kword{spin Hall effect, spin-orbit interaction, antidot, STM,
resonant scattering, spin filter}

\begin{document}
\maketitle


The spin-orbit (SO) interaction in semiconductors
has attracted a lot of attention for its possible application
to manipulate electron spins in spin-based electronics,
``spintronics.''\cite{Zutic}
The SO interaction is a relativistic effect and written as
\begin{equation}
H_{\rm SO} =\frac{\lambda}{\hbar} \bm{\sigma} \cdot
\left[\bm{p} \times \bm{\nabla} V(\bm{r}) \right]
\label{eq:SOorg}
\end{equation}
for electrons in the vacuum, where $V(\bm{r})$ is
an external potential and $\bm{\sigma}$
indicates the electron spin $\bm{s}=\bm{\sigma}/2$.
The coupling constant is given by
$\lambda=-\hbar^2/(4 m_0^2 c^2)$ with
electron mass $m_0$ and velocity of light $c$.
For conduction electrons in direct-gap semiconductors,
the SO interaction is expressed in the same form.
The coupling constant $\lambda$ is much larger than
the value in the vacuum owing to the band effect,
particularly in narrow-gap semiconductors such as InAs
\cite{Winkler}.

A well-known example is the Rashba SO interaction
in two-dimensional electron gas (2DEG) in semiconductor
heterostructures.\cite{Rashba,Rashba2} An electric field
perpendicular to the 2DEG in the $xy$ plane,
$V(\bm{r})=e {\cal E} z$, gives rise to
\begin{equation}
H_{\rm SO} =\frac{\alpha}{\hbar} (p_y\sigma_x-p_x\sigma_y),
\label{eq:Rashba}
\end{equation}
with $\alpha=e {\cal E} \lambda$. The large values of $\alpha$
have been reported in experiments.\cite{Nitta,Grundler,Yamada}
The spin transistor proposed by Datta and Das is based on
this Rashba SO interaction because of its tunability by the
external field.\cite{Datta-Das} The electron spins are
manipulated by the SO interaction in semiconductor, which are
injected from a ferromagnet and detected by another ferromagnet.

The SO interaction may also be useful for the spin injection,
instead of using ferromagnets, in the spintronics devices.
The spin Hall effect (SHE) is one of the phenomena to generate
a spin current. The effect is categorized into two, intrinsic
and extrinsic SHEs. The former is induced by the
drift motion of holes in the SO-split valence
bands\cite{Murakami,Wunderlich} or that of electrons in the
conduction band in the presence of Rashba SO
interaction.\cite{Sinova} The latter stems from the impurity
scattering. For centrally symmetric potential around an
impurity, $V(r)$, eq.\ (\ref{eq:SOorg}) is rewritten as
\begin{equation}
H_{\rm SO} =-\lambda\frac{2}{r} \frac{dV}{dr}
\bm{l} \cdot \bm{s},
\label{eq:SO3D}
\end{equation}
where $\bm{l}=(\bm{r} \times \bm{p})/\hbar$ is the angular momentum.
This results in the skew scattering: Accompanied by the
scattering from direction $\bm{n}$ to $\bm{n'}$, the spin is
polarized in $(\bm{n} \times \bm{n'})/
|\bm{n} \times \bm{n'}|$.\cite{Mott,Landau}
In an optical experiment of Kerr rotation, Kato {\it et al}.\
have observed a spin accumulation at sample edges transverse
to the electric current in $n$-type GaAs.\cite{Kato}
This is ascribable to the extrinsic SHE due to the
scattering of conduction electrons by the screened Coulomb
potential around charged impurities.\cite{Engel}

In the present letter, we focus on the extrinsic SHE in 2DEG
in semiconductor heterostructures. We begin with the
quantum mechanical formulation of the effect. Although
the extrinsic SHE is usually described by a semi-classical
theory considering the skew scattering and ``side jump''
effects,\cite{Engel} the quantum theory is required to
fully understand the SHE and should be useful in designing
spintronics devices based on 2DEG.
We stress that the SHE is easier to understand in 2DEG than
in three-dimensional case.
Second, we examine the SHE in 2DEG by an artificial potential
created by antidot, STM tip, etc. The antidot is a small
metallic electrode fabricated above the 2DEG, as
schematically shown in the inset in Fig.\ 1, to create
a scattering potential for electrons.
The potential is electrically tunable and may be
attractive as well as repulsive. We show that
the SHE is significantly enhanced by the resonant scattering
when the attractive potential is properly tuned. Finally, we
propose a three-terminal device including an antidot. Until
now, several spin-filtering devices have been proposed using
semiconductor nanostructures with SO
interaction.\cite{Kiselev,Koga,Streda,Pareek,Hankiewicz,Ohe,me}
Recently, Yamamoto and Kramer have studied a three-terminal spin
filter with a repulsive antidot potential.\cite{Yamamoto}
We show that a similar device with an attractive antidot potential
could be a spin filter with an efficiency of more than 50\%
by the tuning to the resonance.


We consider a scattering problem of an electron in the $xy$ plane
by an axially symmetric potential $V(r)$ ($r=\sqrt{x^2+y^2}$).
The SO interaction is given by
\begin{equation}
H_{\rm SO} =-\lambda \frac{2}{r} \frac{dV}{dr} l_z s_z
\equiv V_1(r) l_z s_z,
\label{eq:SO2D}
\end{equation}
with $l_z$ and $s_z$ being the $z$ component of angular momentum
and spin operators. $V_1(r)=-(2 \lambda/r)V'(r)$ has the
same sign as $V(r)$ when $|V(r)|$ is a monotonically decreasing
function of $r$ and $\lambda>0$. Assuming that $V(r)$ is
smooth in the scale of lattice constant, we adopt the
effective mass equation
\begin{equation}
\left[ -\frac{\hbar^2}{2m^*} \Delta + V(r) + V_1(r)l_z s_z \right]
\psi({\bf r}) =E \psi({\bf r}),
\label{eq:Schroedinger}
\end{equation}
for an envelope function $\psi({\bf r})$ with
effective mass $m^*$. The Dresselhaus SO interaction is neglected,
which stems from an inversion asymmetry of the crystal.\cite{Dresselhaus}

Note that $l_z$ and $s_z$ are conserved in
eq.\ (\ref{eq:Schroedinger}), in contrast to the three-dimensional
case with eq.\ (\ref{eq:SO3D}), which simplifies the discussion.
For $s_z=\pm 1/2$, an electron feels the potential of
$V(r) \pm V_1(r) l_z/2$. In consequence the scattering
for components of $l_z>0$ ($l_z<0$) is enhanced (suppressed)
by the SO interaction for $s_z=1/2$ when $V_1(r)$ has the same sign
as $V(r)$. The effect is opposite for $s_z=-1/2$. This is the origin
of the extrinsic SHE.

We adopt a partial wave expansion for the scattering problem with
$l_z=m=0$, $\pm 1$, $\pm 2$, $\cdots$.\cite{Aharonov-Bohm}
As an incident wave, we consider a plane wave propagating in $x$
direction, $e^{ikx}$, with spin $s_z=1/2$ or $-1/2$.
$E=\hbar^2 k^2/(2m^*)$. The plane wave is expanded as
\begin{equation}
e^{ikx}=e^{ikr \cos \theta}=\sum_{m=-\infty}^{\infty}
i^m J_{m}(kr) e^{im\theta},
\label{eq:incident}
\end{equation}
where $\theta$ is the angle from $x$ direction and
$J_{m}$ is the $m$th Bessel function. Its asymptotic form at
$r \rightarrow \infty$ is given by
$J_{m}(kr)\sim \sqrt{2/(\pi kr)} \cos(kr-m\pi/2-\pi/4)$.
In the solution of eq.\ (\ref{eq:Schroedinger}),
$J_{m}(kr)$ in
eq.\ (\ref{eq:incident}) is replaced by $R_m^{\pm}(r)$ for
$s_z=\pm 1/2$,\cite{com1} which satisfies
\[
\left[
-\frac{\hbar^2}{2m^*} \left(
\frac{d^2}{dr^2}+\frac{1}{r}\frac{d}{dr}-\frac{m^2}{r^2}
\right)+V(r) \pm \frac{m}{2} V_1(r)
\right] R_m^{\pm}(r) 
\]
\begin{equation}
= E R_m^{\pm}(r).
\label{eq:Req}
\end{equation}
Its asymptotic form determines the phase shift $\delta_m^{\pm}$:
\begin{equation}
R_m^{\pm}(r) \sim \sqrt{\frac{2}{\pi kr}} e^{i \delta_m^{\pm}}
\cos \left( kr-\frac{m\pi}{2}-\frac{\pi}{4}+\delta_m^{\pm}
\right).
\label{eq:Req2}
\end{equation}
From eqs.\ (\ref{eq:Req}) and (\ref{eq:Req2}),
we immediately obtain the relations of
$\delta_{-m}^{\pm}=\delta_m^{\mp}$, indicating the time reversal
symmetry. The SO interaction does not work for the $S$ wave
($m=0$): $\delta_0^{+}=\delta_0^{-} \equiv \delta_0$.

The scattering amplitude $f^{\pm}(\theta)$ for $s_z=\pm 1/2$
is expressed in terms of the phase shifts:
$f^{\pm}(\theta) = f_1(\theta) \pm f_2(\theta)$,
\begin{eqnarray}
f_1(\theta) &=&
\frac{1}{i \sqrt{2\pi k}} \biggl[
e^{2i\delta_0}-1
\nonumber \\
& & +
\sum_{m=1}^{\infty}
\left(e^{2i\delta_m^+} + e^{2i\delta_m^-}-2 \right)
\cos m\theta \biggr],
\\
f_2(\theta) &=&
\frac{1}{\sqrt{2\pi k}} \sum_{m=1}^{\infty}
(e^{2i\delta_m^+}-e^{2i\delta_m^-})
\sin m\theta.
\end{eqnarray}
The scattering cross section is given by
$\sigma^{\pm}(\theta)=|f^{\pm}(\theta)|^2$. Hence
the spin polarization of the scattered wave in $\theta$
direction is expressed as
\begin{equation}
P_z=\frac{|f^+|^2-|f^-|^2}{|f^+|^2+|f^-|^2}=
\frac{2{\rm Re}(f_1 f_2^*)}{|f_1|^2+|f_2|^2},
\label{eq:eSHEpol}
\end{equation}
when the incident electron is unpolarized.
This formula is analogous to that of skew scattering in
three-dimensions,\cite{Mott,Landau} and one of the main results
in the present letter. The spin is polarized in $z$ direction
and $P_z(-\theta)=-P_z(\theta)$.


\begin{figure}
\begin{center}
\includegraphics[width=7cm]{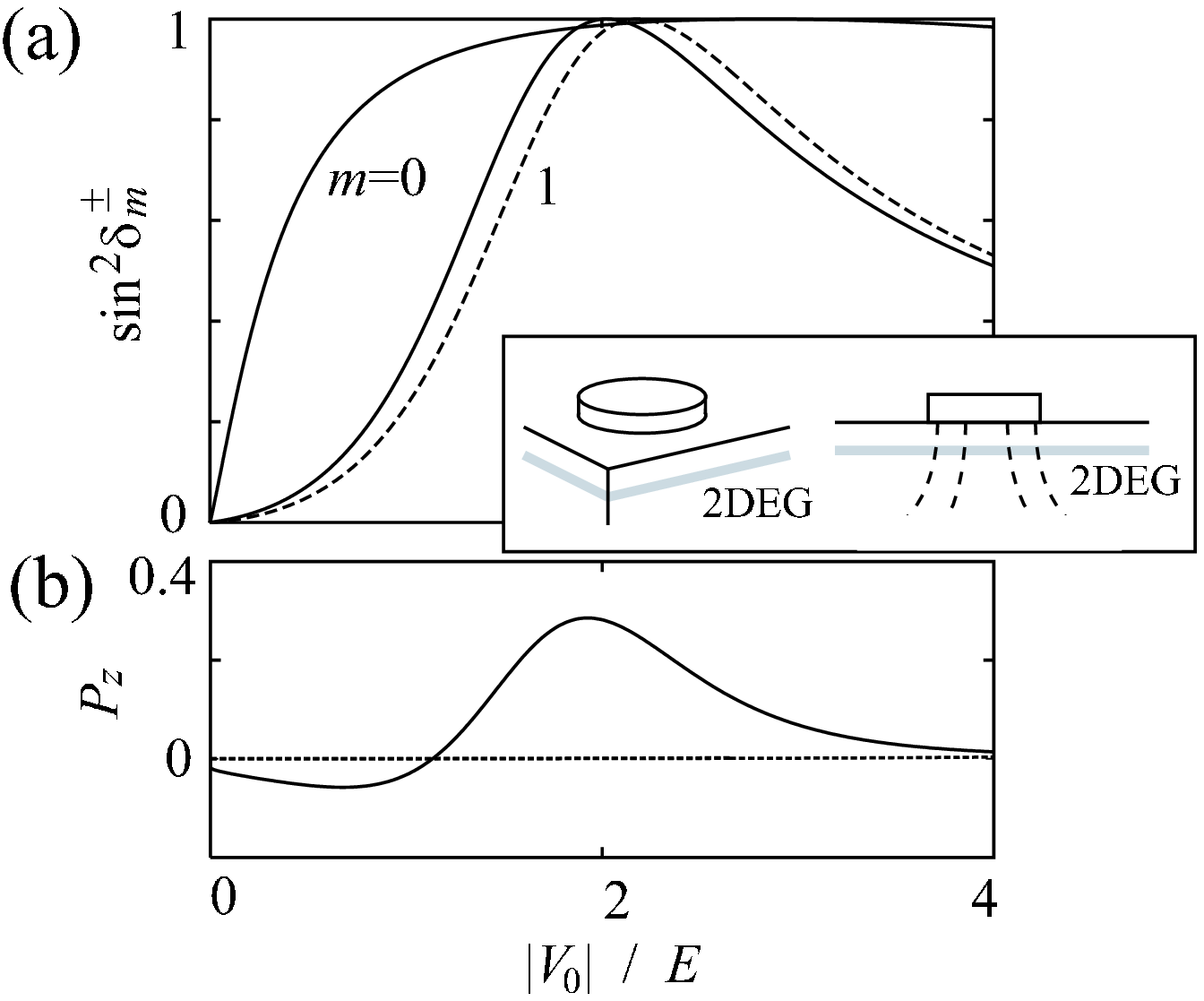}
\end{center}
\caption{Extrinsic SHE due to the scattering by
a potential well, $V(r)=V_0\theta(a-r)$ ($V_0<0$), for 2DEG.
$a=1/k$. The strength of the SO interaction is
$\lambda k^2=0.01$ ($\lambda=117.1$\AA$^2$, $2\pi/k=70$nm).
(a) The scattering probability of each partial wave,
$\sin^2 \delta_m^{\pm}$, and
(b) spin polarization $P_z$ at $\theta =-\pi/2$,
as functions of the potential depth $|V_0|$ [normalized
by electron energy $E=\hbar^2 k^2/(2m^*)$]. In (a),
solid and broken lines indicate the cases of
$s_z=1/2$ and $-1/2$, respectively, for $m=1$
($\delta_{-1}^{\pm}=\delta_1^{\mp}$). The scattering
probability for $|m| \ge 2$ is negligibly small.
Inset: schematic drawing of artificial potential on 2DEG
created by antidot. The potential is attractive (repulsive)
when a positive (negative) voltage is applied to the antidot.
}
\label{f1}
\end{figure}

Now we examine the SHE due to the scattering by an attractive
potential. The simplest example is a potential well,
$V(r)=V_0\theta(a-r)$ ($V_0<0$, $a>0$),
where $\theta(t)$ is a step function [$\theta(t)=1$ for
$t>0$, $0$ for $t<0$]. Then $V_1=(2\lambda/a) V_0 \delta(r-a)$
with $\delta$-function $\delta(t)$.
The phase shifts $\delta_m^{\pm}$ are calculated by solving
eq.\ (\ref{eq:Req}).

Figure 1 shows the scattering probability of each partial wave,
$\sin^2 \delta_m^{\pm}$, and
spin polarization $P_z$ at $\theta=-\pi/2$, as functions of
the potential depth $|V_0|$. The strength of the SO interaction
is set to be $\lambda k^2=0.01$, which corresponds to the
value of InAs, $\lambda=117.1$\AA$^2$,\cite{Winkler} with the
electron wavelength $2\pi/k=70$nm.
The radius of the potential well is $a=1/k$.
With an increase in $|V_0|$, the scattering probability
increases and becomes unity at some
values of $|V_0|$ (unitary limit with $\delta_m^{\pm}=\pi/2$) for
$m=0$ ($S$ wave) and $m=\pm 1$ ($P$ wave). This is due to the
resonant scattering through virtual bound states in the potential
well. The resonant width is narrower for larger $|m|$ because
of the centrifugal potential $\propto m^2/r^2$ separating the
bound states from the outer region.
Around the resonance of the $P$ waves, a difference in
$\delta_1^{+}-\delta_1^{-} \equiv \Delta \delta_1$ results in
a large spin-polarization $P_z \approx 30\%$. Around the resonance,
$(\delta_1^{+}+\delta_1^{-})/2 \approx \pi/2$,
formula (\ref{eq:eSHEpol}) yields
\[
P_z(\theta=-\pi/2) \approx
\frac{2 \sin^2 \delta_0 \sin \Delta \delta_1}
{\sin^2 \delta_0 + \sin^2 \Delta \delta_1}
\]
when $\delta_m^{\pm}$ ($|m| \ge 2$) is negligibly small.

\begin{figure}
\begin{center}
\includegraphics[width=6cm]{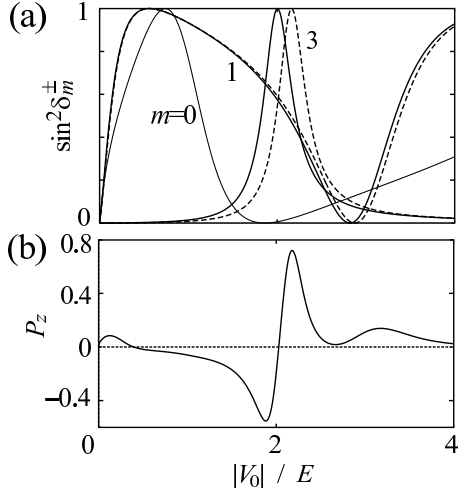}
\end{center}
\caption{Extrinsic SHE due to the scattering by
a potential well, $V(r)=V_0\theta(a-r)$ ($V_0<0$), for 2DEG.
$a=2/k$. The strength of the SO interaction is
$\lambda k^2=0.01$.
(a) The scattering probability of each partial wave,
$\sin^2 \delta_m^{\pm}$, and
(b) spin polarization $P_z$ at $\theta =-\pi/2$,
as functions of the potential depth $|V_0|$ [normalized
by electron energy $E=\hbar^2 k^2/(2m^*)$]. In (a),
solid and broken lines indicate the cases of
$s_z=1/2$ and $-1/2$, respectively, for $m>0$
($\delta_{-m}^{\pm}=\delta_m^{\mp}$).
The data for $|m|=2$ are omitted.
}
\label{f2}
\end{figure}

Figure 2 shows the calculated results for a wider potential well,
$a=2/k$. The resonant scattering takes place for $0 \le |m| \le 3$
(not shown for $|m|=2$ because the resonance with even $m$ is
not relevant to the spin polarization at $\theta=-\pi/2$).
Around the resonance of $F$ waves ($|m|=3$), 
$P_z$ is enhanced to 72\%. In general,
a sharper resonance enlarges
$\delta_{m}^{+}-\delta_{m}^{-}$
for larger $|m|$,
which results in a larger polarization.

The extrinsic SHE is expected even with a repulsive potential
in the presence of SO interaction, eq.\ (\ref{eq:SO2D}).
We solve the scattering problem with a potential barrier,
$V(r)=V_0\theta(a-r)$ with $V_0>0$ and $a=2/k$.
We find that
the spin polarization $P_z(\theta=-\pi/2)$ is less than
0.5\% in the range of $0<V_0/E<4$ (not shown),
which is much smaller than
the values in Fig.\ 2 with attractive potential. This indicates
the importance of the resonant scattering for the enhancement
of SHE.



\begin{figure}
\begin{center}
\includegraphics[width=6cm]{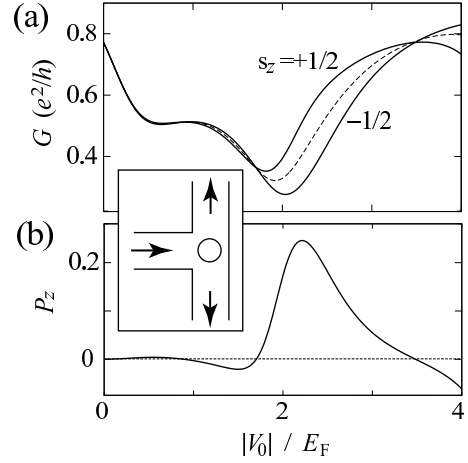}
\end{center}
\caption{Numerical results of spin injection
in a three-terminal device including a tunable antidot potential,
schematically shown in the inset.
(a) Conductance $G_{\pm}$ from the left lead to the lower
lead with spin $s_z=\pm 1/2$, and (b) spin polarization of
the current in the lower lead, as functions of the depth
of attractive potential $|V_0|$ (normalized
by Fermi energy $E_{\rm F}$). The strength of SO
interaction is $\lambda k_{\rm F}^2=0.028$, with $k_{\rm F}$
being the Fermi wavenumber ($\lambda=117.1$\AA$^2$,
$2\pi/k_{\rm F}=40$nm).
The radius of the potential is $a=2/k_{\rm F}$.
In (a), broken line indicates the
conductance per spin in the absence of SO interaction.
}
\label{f3}
\end{figure}

Making use of a tunable antidot potential, we propose a
three-terminal device as an efficient spin filter
[inset in Fig.\ 3].
Unpolarized electrons are injected from the left lead
(connected to the source electrode)
and outgoing into upper and lower leads (connected to
the drain electrodes). The voltages are equal in the two drains.
We assume a hard-wall potential for the boundaries of leads
and a smooth potential well for an antidot,
$V(r)=V_0$ ($r<a-\Delta/2$), $(V_0/2)[1-\sin[\pi(r-a)/\Delta]]$
($|r-a|<\Delta/2$), $0$ ($r>a+\Delta/2$), where $V_0<0$ and
$r$ is the distance from the center of the junction.
We set $\Delta=0.7a$. In the presence of SO interaction,
eq.\ (\ref{eq:SO2D}), $s_z$ is conserved, whereas
$l_z$ is not owing to the lack of rotational symmetry of
the system.

The conductance $G_{\pm}$ for $s_z=\pm 1/2$ is numerically
evaluated in the same way as in ref.\ 22,
using the Green function's recursion method on the
tight-binding model of a square lattice ($31 \times 31$ sites in
the junction area). The temperature is $T=0$.
The spin polarization is defined as
$P_z =(G_+ -G_-)/(G_+ +G_-)$.

Figure 3 shows the conductance $G_{\pm}$ from the source
to the lower drain, with the spin polarization $P_z$,
as functions of the potential depth $|V_0|$.
(The spin polarization is $-P_z$ in the upper drain.)
We assume that the Fermi wavelength is $2\pi/k_{\rm F}=40$nm
($k_{\rm F}$ is the Fermi wavenumber) and thus
$\lambda k_{\rm F}^2=0.028$ with $\lambda=117.1$\AA$^2$.
The potential radius is $a=2/k_{\rm F}$
and the width of leads is $W=4a=8/k_{\rm F}$. Then
there are two conduction modes in the leads at the
Fermi energy $E_{\rm F}$.
The spin polarization is enhanced to 25\% around
$|V_0|/E_{\rm F}=2$, which is attributable to the resonant
scattering via a virtual bound state around the
antidot,\cite{Yokoyama} as discussed before.

To examine the resonance in detail, we make a channel analysis
for two incident modes from the left lead.
In Fig.\ 4, we plot $P_z$ separately for the lowest mode,
$e^{i k_1 x} \cos(\pi y/W)$, and
for the second mode, $e^{i k_2 x} \sin(2\pi y/W)$
[$k_1^2+(\pi /W)^2=k_2^2+(2\pi /W)^2=k_{\rm F}^2$].
The lowest mode plays a main role in the enhancement of
spin polarization around the resonance. Since
we could  selectively inject the lowest mode to the junction,
e.\ g., using a quantum point contact fabricated on the left lead,
our device could be a spin filter with an efficiency of more
than 50\%.

In conclusion, we have formulated the extrinsic SHE for 2DEG
in semiconductor heterostructures, using the quantum mechanics.
We have examined the SHE due to the scattering by a tunable
potential created by antidot, STM tip, etc. The resonant
scattering significantly enhances the SHE with an attractive
potential. We have proposed a three-terminal device including
an antidot, as an efficient spin filter.

\begin{figure}
\begin{center}
\includegraphics[width=6cm]{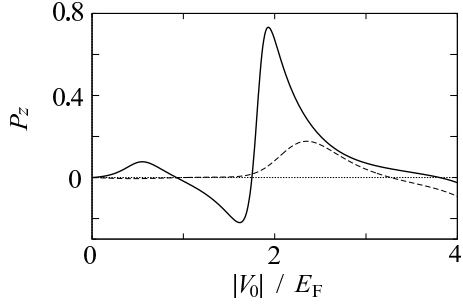}
\end{center}
\caption{Spin polarization $P_z$ for each channel in the
incident current in the three-terminal device given
in Fig.\ 3. Solid and broken lines indicate $P_z$ for
the lowest mode and second mode, respectively.
The parameters are the same as in Fig.\ 3.
}
\label{f4}
\end{figure}

A three-terminal spin filter without antidot has been
studied by Kiselev and Kim in the presence
of Rashba SO interaction.\cite{Kiselev} They have pointed out an
enhancement of spin polarization by the resonant scattering
at the junction
when the Fermi energy of 2DEG is tuned. In their device,
the direction of spin polarization is tilted
from the $z$ direction perpendicular to the plane. In our device,
the spin is polarized in $z$ direction, which is easier to
detect by the optical experiment,\cite{Kato} and above all,
more suitable to the spintronics devices.

The extrinsic SHE enhanced by (many-body) resonant scattering
has been examined for metallic systems with magnetic
impurities.\cite{Fert,Fert2,Guo} In the case of
semiconductor heterostructures, however, we have a great advantage
in the tunability of potential. The SHE by the resonant scattering
at a single potential can be investigated in details.

We make some comments regarding our device.
(i) The electron-electron interaction has been neglected in
our calculations. The number of electrons trapped in the potential
well is given by the Friedel sum rule,
$(1/\pi)\sum_m \sum_{\sigma} \delta_m^{\sigma}$. The Hartree
potential from the electrons should be considered
although the Coulomb blockade is irrelevant to the antidot potential
without tunnel barriers in contrast to usual quantum dots.
Therefore, the values of $|V_0|$ at the resonance have been
underestimated.
(ii) It is required to create such a deep potential as
$|V_0| \sim E_{\rm F}$ in designing the device.
(iii) We have assumed that the antidot potential $V(\bm{r})$ does
not depend on $z$. Otherwise, the Rashba-type SO interaction, eq.\
(\ref{eq:Rashba}) with $\alpha=\lambda(\partial V/\partial z)$,
has to be added to eq.\ (\ref{eq:SO2D}).
It would make an effective magnetic field in the $xy$ plane and thus
decrease the spin polarization in $z$ direction.

\section*{Acknowledgment}
This work was partly supported by the Strategic Information
and Communications R\&D Promotion Program (SCOPE) from the
Ministry of Internal Affairs and Communications of Japan, and
by Grant-in-Aid for Scientific Research from
the Japan Society for the Promotion of Science.


\begin{thebibliography}{99}
\bibitem{com1}
The asymptotic form of $\psi(\bm{r})$ is given by
$\psi \sim e^{ikx} + f^{\pm}(\theta) e^{i(kr+\pi/4)}/\sqrt{r}$
for $s_z=\pm 1/2$ [phase $\pi/4$ on the exponent ensures the
optical theorem that the total cross section is identical to
$\sqrt{8\pi/k} {\rm Im} f^{\pm}(0)$].
The scattering amplitude
$f^{\pm}(\theta)=\sum_m f_m^{\pm}e^{im\theta}$ is
related to the S-matrix by
$S_m^{\pm}=1+i\sqrt{2\pi k}f_m^{\pm}=e^{2i\delta_m^{\pm}}$
with $\delta_m^{\pm}$ being the phase shifts.
\end{thebibliography}
\end{document}